\documentclass[aps,prl,reprint,superscriptaddress,amsmath,amssymb]{revtex4-1}
\usepackage{bm}
\usepackage[pdftex]{graphicx,hyperref}
\hypersetup{colorlinks=true, anchorcolor=blue, linkcolor=blue, citecolor=blue, filecolor=blue, urlcolor=blue}

\begin{document}

\title{Fully Coupled Two-Fluid Dynamics in Superfluid $^4$He: Anomalous Anisotropic Velocity Fluctuations in Counterflow}
\author{Satoshi Yui}
\affiliation{Research and Education Center for Natural Sciences, Keio University, 4-1-1 Hiyoshi, Kohoku-ku, Yokohama 223-8521, Japan}

\author{Hiromichi Kobayashi}
\affiliation{Research and Education Center for Natural Sciences, Keio University, 4-1-1 Hiyoshi, Kohoku-ku, Yokohama 223-8521, Japan}
\affiliation{Department of Physics, Hiyoshi Campus, Keio University, 4-1-1 Hiyoshi, Kohoku-ku, Yokohama 223-8521, Japan}

\author{Makoto Tsubota}
\affiliation{Department of Physics \& Nambu Yoichiro Institute of Theoretical and Experimental Physics (NITEP) \& The OCU Advanced Research Institute for Natural Science and Technology (OCARINA), Osaka City University, 3-3-138 Sugimoto, Sumiyoshi-ku, Osaka 558-8585, Japan}

\author{Wei Guo}
\affiliation{National High Magnetic Field Laboratory, 1800 East Paul Dirac Drive, Tallahassee, Florida 32310, USA}
\affiliation{Mechanical Engineering Department, Florida State University, Tallahassee, Florida 32310, USA}

\date{\today}

\begin{abstract}
We investigate the thermal counterflow of the superfluid $^4$He by numerically simulating three-dimensional fully coupled dynamics of the two fluids, namely quantized vortices and a normal fluid.
We analyze the velocity fluctuations of the laminar normal fluid arising from the mutual friction with the quantum turbulence of the superfluid component.
The streamwise fluctuations exhibit higher intensity and longer-range autocorrelation, as compared to transverse ones.
The anomalous fluctuations are consistent with visualization experiments [Mastracci {\it et al.}, Phys. Rev. Fluids {\bf 4}, 083305 (2019)], and our results confirm their analysis with simple models on the anisotropic fluctuations.
This success validates the model of the fully coupled dynamics and paves the way for solving some outstanding problems in this two-fluid system.
\end{abstract}

\maketitle

%%%
{\it Introduction.}---Quantum turbulence (QT) refers to the turbulent flow in a superfluid \cite{vinen02,halperin09,tsubota13,nemirovskii13,barenghi14,tsubota17}, which can occur in a wide range of coherent matter-wave systems, {\it e.g.}, superfluid $^3$He and $^4$He \cite{vinen06}, atomic Bose--Einstein condensates (BECs) \cite{henn09}, neutron stars \cite{packard72}, and galactic dark-matter BECs \cite{sikivie09}.
At finite temperatures, the interaction between QT and the thermal component can lead to intriguing hydrodynamical behaviors that are new to physics.
In this study, we address an outstanding phenomenon of the coupled dynamics in the superfluid $^4$He, {\it i.e.}, the velocity fluctuations of the thermal component caused by QT.

Liquid $^4$He exhibits superfluidity below $T_c = 2.17 ~\mathrm{K}$ \cite{kapitza38,tilley90,donnelly91}.
Superfluid $^4$He (He II) can be understood via the two-fluid model \cite{tisza38,landau41}.
In this model, He II is described by a mixture of an inviscid superfluid and a viscous normal fluid (thermal excitations).
The ratio of superfluid density $\rho_s$ to the normal-fluid density $\rho_n$ depends on temperature.
The normal fluid and superfluid exhibit individual velocities ${\bm v}_n$ and ${\bm v}_s$, respectively. 
In the superfluid component, a quantized vortex appears as rotational motion, which exhibits quantum circulation $\kappa = 1.0 \times 10^{-3} ~\mathrm{cm^2/s}$.
The angstrom-sized vortex core can be considered as the filament with $\kappa$, which is termed as the vortex filament model (VFM).
Conversely, the normal-fluid component behaves in a manner similar to a viscous classical fluid.
The quantized vortices and the normal fluid affect each other via mutual friction (MF), and coupled dynamics is essentially important to understand He II.

QT is a tangle of quantized vortices, and this tangle produces a turbulent velocity field of the superfluid.
The typical experiment to generate QT corresponds to a thermal counterflow \cite{tough82}, which is a relative flow of the two fluids.
In a closed channel, the temperature gradient is applied via a heater.
The normal fluid flows from the heater to the cooler side to transfer heat.
The superfluid flows to the heater to satisfy the mass conservation $\int_{\mathcal S} (\rho_n {\bm v}_n + \rho_s {\bm v}_s) dS = {\bm 0}$, where the integral is performed over the channel cross section.
When the relative velocity ${\bm v}_{ns} = {\bm v}_n - {\bm v}_s$ exceeds a critical value, QT appears in the thermal counterflow.
A vortex line density $L = \frac{1}{\Omega} \int_{\mathcal L} d\xi$ is measured in a statistically steady state with the sample volume $\Omega$, the integral path ${\mathcal L}$ along the vortex filaments, and the arc length $\xi$ along the filaments.
The value of $L$ increases with the mean relative velocity $V_{ns} = | \langle{\bm v}_{ns} \rangle |$ with spatial average $\langle \cdot \rangle$ and obeys the steady-state relation 
\begin{equation}
  L^{\frac{1}{2}} = \gamma (V_{ns} - V_0)
  \label{eq:vinen}
\end{equation}
based on Vinen's equation employing the temperature-dependent parameter $\gamma$ and a fitting parameter $V_0$ \cite{vinen57c,tough82}.

Extensive experimental studies by Tough {\it et al.} revealed that there are two turbulent regimes in counterflow:
a T1 state characterized by smaller values of $\gamma$ and a T2 state with larger $\gamma$ \cite{tough82}.
They suggested that the T1 state is associated with turbulence only in the superfluid while in the T2 state both fluids are likely turbulent.
Melotte and Barenghi \cite{melotte98} performed linear stability analysis of the normal fluid in the T1 state and suggested that the laminar normal fluid could become unstable due to MF.
Experimental confirmation of the doubly turbulent T2 state in counterflow was first provided by Guo {\it et al.} \cite{guo10}.
More detailed subsequent studies revealed a nonclassical energy spectrum and exceptionally high turbulence intensity in the T2 state \cite{marakov15}, the understanding of which is a topic of current interests \cite{gao17,bao18,biferale19}.

This Letter is concerned with some striking new observations from a more recent flow visualization experiment on counterflow turbulence conducted by Mastracci and Guo \cite{mastracci18}.
In their particle tracking velocimetry (PTV) measurement, they showed that in the T1 state, there exist unexpected anisotropic velocity fluctuations in the \emph{laminar} normal fluid.
Inspired by early analysis and simulations \cite{vinen57c,idowu00,kivotides00}, Mastracci {\it et al.} suggested that these fluctuations may arise due to the MF drag in the normal fluid from individual quantized vortices \cite{mastracci19}, and they supported this suggestion by analyzing various simple models.
However, a more detailed understanding of the observations is possible only with the fully coupled two-fluid dynamics.

Two major methods address the three-dimensional coupled dynamics in He II.
The first method is to use the Hall--Vinen--Bekarevich--Khalatnikov (HVBK) equations for both fluids, where quantized vortices are coarse-grained \cite{donnelly91,bertolaccini17,kobayashi19,biferale19}.
The model is useful in studying properties larger than the mean inter-vortex spacing $\ell$ of QT.
However, the model does not describe the dynamics of quantized vortices although it is essential for QT.
The other method is to employ the VFM for the superfluid coupled with the HVBK equations for the normal fluid \cite{kivotides00,kivotides07,kivotides11,khomenko17,yui18}.
Recently, Yui {\it et al.} demonstrated the calculation involving QT in their study of the normal-fluid velocity profile in counterflow \cite{yui18}.
Nevertheless, a coarse-grained MF was used in that work, which obscures any possible normal-fluid vortices near the vortex filaments.

In this Letter, we investigate how the laminar normal fluid is disturbed by QT through the MF in the T1 state.
We introduce a numerical framework based on the VFM for the superfluid coupled with HVBK equations for the normal fluid without any spatial coarse-graining of the MF.
The three-dimensional simulation based on this model allows us to resolve eddy structures generated by the MF in the laminar normal fluid in the vicinity of the vortex tangle.
The calculated streamwise velocity fluctuations in the normal fluid exhibit higher intensity and a longer autocorrelation range.
The results are consistent with the PTV observations, and confirm their explanation with simple models \cite{mastracci19}.
This work not only elucidates the origin of the intriguing velocity fluctuations in laminar normal fluid but also validates this model of the coupled dynamics, making it a valuable tool for solving various unsolved problems in this two-fluid system.

%%%
{\it Coupled dynamics of quantized vortices and normal fluid.}---The VFM is used as one of the most powerful tools to describe the dynamics of quantized vortices  \cite{schwarz85,schwarz88,adachi10,baggaley13,kondaurova14,baggaley15,khomenko15,yui15,gao18,tsubota17}.
The position vector ${\bm s}$ of the filaments are represented by the parametric form ${\bm s} = {\bm s}(\xi)$ with arc length $\xi$.
The superfluid velocity is obtained by the Biot--Savart integral as follows:
$
  {\bm v}_s ({\bm r}) = \frac{\kappa}{4 \pi} \int_{\mathcal L} \frac{({\bm s}_1 - {\bm r})\times d{\bm s}_1}{|{\bm s}_1 - {\bm r}|^3} + {\bm v}_{s,b} + {\bm v}_{s,a}.
$
Specifically, ${\bm v}_{s,b}$ is a velocity induced for boundary condition, and ${\bm v}_{\rm s,a}$ is an externally applied velocity.
We employ the full Biot--Savart integral containing the non-local interactions \cite{adachi10}.
Eventually, the velocity of the filaments is as follows \cite{barenghi83,schwarz85}:
\begin{equation}
  \frac{d\bm s}{dt} = {\bm v}_s + \alpha {\bm s}' \times {\bm v}_{ns} - \alpha' {\bm s}' \times \left( {\bm s}' \times {\bm v}_{ns} \right),
  \label{eq:filament}
\end{equation}
where ${\bm s}'$ denotes the unit tangent vector of the filaments.
The terms including temperature-dependent coefficients $\alpha$ and $\alpha'$ show the MF with the normal fluid.

The dynamics of the normal fluid is given by the HVBK equations \cite{barenghi83,donnelly91}:
\begin{equation}
  \frac{\partial {\bm v}_n}{\partial t} + ({\bm v}_n \cdot \nabla) {\bm v}_n = - \frac{1}{\rho} \nabla P + \nu_n \nabla^2 {\bm v}_n + \frac{1}{\rho_n} {\bm F}_{ns},
  \label{eq:navier}
\end{equation}
by using the kinetic viscosity $\nu_n = \eta_n / \rho_n$ of the normal fluid and the effective pressure gradient $\nabla P$. 
Here, the MF force
$
  {\bm F}_{ns} ({\bm r}) = \frac{1}{\Omega' ({\bm r})} \int_{{\mathcal L} ' ({\bm r})} {\bm f}(\xi) d\xi
$
is obtained by the integral of the MF ${\bm f}$ per unit length of the filaments:
$
  {\bm f}(\xi) / \rho_s \kappa = \alpha {\bm s}' \times \left ( {\bm s}' \times {\bm v}_{ns} \right ) + \alpha' {\bm s}' \times {\bm v}_{ns}$.
${\mathcal L}'({\bm r})$ denotes the filaments in the local sub-volume $\Omega ' ({\bm r})$ at the position ${\bm r}$.
The size of $\Omega '$ determines the coupling length scale (See Supplements).
In the study, we employ the local coupling condition $\ell^3 > \Omega'$, {\it i.e.}, the MF ${\bm F}_{ns}$ only affects the normal fluid at the position of the vortex filaments in contrast to a preceding study \cite{yui18}.
We use the incompressible condition $\nabla \cdot {\bm v}_n = 0$ as a closure.

%%%
{\it Numerical simulation.}---We perform numerical simulations of the coupled dynamics in thermal counterflow.
First, we check the relation of Eq. (\ref{eq:vinen}) and velocity profiles of the two fluids to know the state of QT.
Second, we examine the three-dimensional structures of the quantized vortices and normal-fluid flow.
Finally, the velocity fluctuations of the normal fluid are statistically analyzed in terms of intensity and autocorrelation.

\begin{figure}
  \centering
  \includegraphics[width=1.0\linewidth]{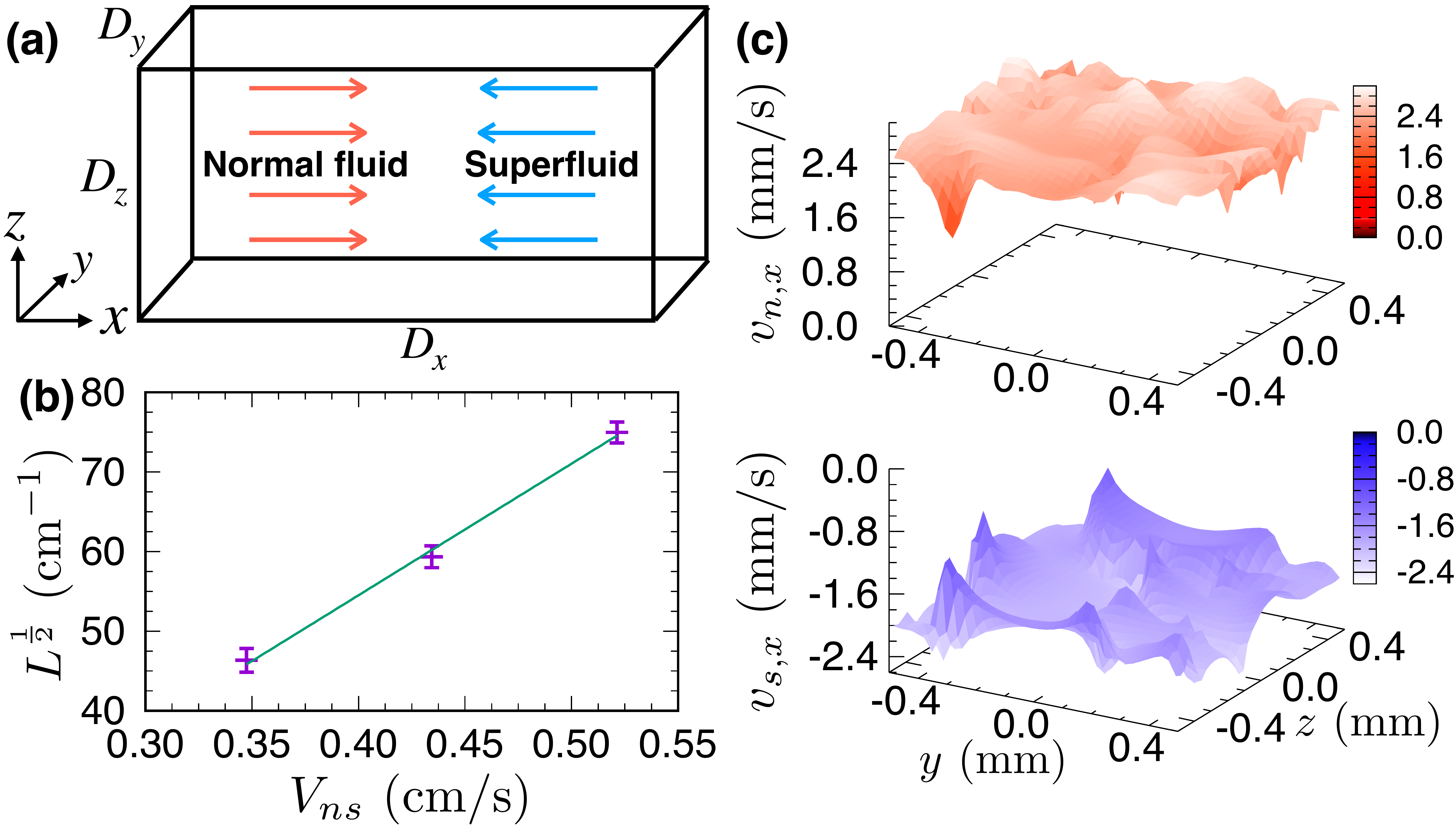}
  \caption
  {
  (a) Schematics of counterflow simulation.
  (b) Averaged values of the vortex line density as a function of the mean relative velocity $V_{ns}$.
  The slope parameter is $\gamma = 165 \pm 9 ~\mathrm{s/cm^2}$.
  (c) Normal-fluid velocity $v_{n,x}$ and superfluid velocity $v_{s,x}$ over the channel cross section in the statistically steady state at $V_n = 2.5 ~\mathrm{mm/s}$.
  \label{density_velocity.pdf}
  }
\end{figure}

The numerical simulations are performed as follows.
The volume of the computational box is $\Omega = D_x D_y D_z = 2.0 ~\mathrm{mm} \times 1.0 ~\mathrm{mm} \times 1.0 ~\mathrm{mm}$, as shown in Fig. \ref{density_velocity.pdf}(a).
The vortex filaments are discretized into a series of points with the separation $\Delta \xi_{\min} = 0.008 ~\mathrm{mm} < \Delta \xi < 0.024 ~\mathrm{mm}$
\footnote
{
The value of $L$ does not change even with the larger spatial resolutions $\Delta \xi_{\min} = 0.016 ~\mathrm{mm}$ and $\Delta x = \Delta y = \Delta z = 0.050 ~{\rm mm}$.
}.
The time development of Eq. (\ref{eq:filament}) is achieved via the 4th order Runge--Kutta method.
When the two filaments approach more closely than $\Delta \xi_{\rm min}$, the filaments are artificially reconnected to each other \cite{schwarz88,adachi10}.
The short filaments with length less than $5 \times \Delta \xi_{\rm min}$ are removed \cite{tsubota00}.
The normal fluid is discretized via the homogeneous spatial grid $N_x N_y N_z = 80 \times 40 \times 40$: the spatial resolutions are $\Delta x = \Delta y = \Delta z = 0.025 ~\mathrm{mm}$.
The sub-volume of the MF is $\Omega' = \Delta x \Delta y \Delta z$.
The time integration of Eq. (\ref{eq:navier}) is achieved by the 2nd order Adams--Bashforth method, and the 2nd order finite-difference method is adopted for spatial differentiation.
Both fluids flow along the $x$-axis.
The periodic boundary condition is applied in all directions.
The initial states correspond to $16$ randomly oriented rings of the quantized vortices and uniform flow of the normal fluid.
The mean velocity of the normal fluid is prescribed as $V_n = | \langle {\bm v}_n \rangle | = 2.0, 2.5, 3.0 ~\mathrm{mm/s}$.
We use ${\bm v}_{s,a} = - (\rho_n / \rho_s) \langle {\bm v}_n \rangle$ as the counterflow condition.
The simulation is performed until $t = 10.0 ~\mathrm{s}$ at $T=1.9 ~\mathrm{K}$.
Temporal-mean values are obtained by averaging values over $5.0 ~\mathrm{s} \leq t \leq 10.0 ~\mathrm{s}$ in statistically steady states.

We obtained the statistically steady state of the two fluids in the counterflow.
The vortex line density $L$ increases from the initial value and fluctuates around some constant values for different $V_n$ (See Supplements).
Thus, QT is in the statistically steady state, where the generation and dissipation of the vortex filaments are balanced.
Figure \ref{density_velocity.pdf}(b) shows the values of $L$ temporally averaged over steady states.
The error bars denote standard deviations.
The mean vortex-line spacing $\ell \sim L^{-\frac{1}{2}}$ is $0.1 ~\mathrm{mm} \lesssim\ell \lesssim 0.2 ~\mathrm{mm}$.
The vortex tangle obeys Eq. (\ref{eq:vinen}), and the coefficient $\gamma = 165 \pm 9 ~\mathrm{s/cm^2}$ exceeds $\gamma_1 \sim 130 ~ \mathrm{s/cm^2}$ of T1 in experiments \cite{childers76,tough82}, but it is still significantly lower than $\gamma_2 \sim 250 ~\mathrm{s/cm^2}$ of T2 \cite{martin83}.
The difference from the observed $\gamma_1$ is potentially because the simulation does not contain the solid channel walls, which can reduce $\gamma$ \cite{baggaley15,yui15}.
Additionally, our value of $\gamma$ is close to the values of the simulations with prescribed uniform flow of normal fluid \cite{adachi10,kondaurova14}.
This implies that the velocity fluctuations of the laminar normal fluid do not significantly amplify $\gamma$.
Figure \ref{density_velocity.pdf}(c) shows snapshots of the velocity profiles over the channel cross section in the steady state at $V_n = 2.5 ~\mathrm{mm/s}$
\footnote
{
The superfluid velocity ${\bm v}_s$ is obtained by the Biot--Savart integral.
In the calculation in Fig. \ref{density_velocity.pdf}(c), we omit the line elements closer to grid points than $\Delta \xi_{\rm min}$.
}.
Specifically, $v_{n,x}$ and $v_{s,x}$ denote the $x$-component of ${\bm v}_n$ and ${\bm v}_s$, respectively.
The profile of $v_{n,x}$ is slightly disturbed while that of $v_{s,x}$ significantly fluctuates.
A Reynolds number $Re_{L} = \Delta v_n I / \nu_n$ is $10^{0}$, where $I = 10^{-1} ~\mathrm{mm}$ denotes integral length and $\Delta v_n$ denotes the fluctuation velocity of the normal fluid, so that the normal fluid should be laminar in the large scales.
The results indicate that QT is in the T1 state.

\begin{figure}
  \centering
  \includegraphics[width=1.0\linewidth]{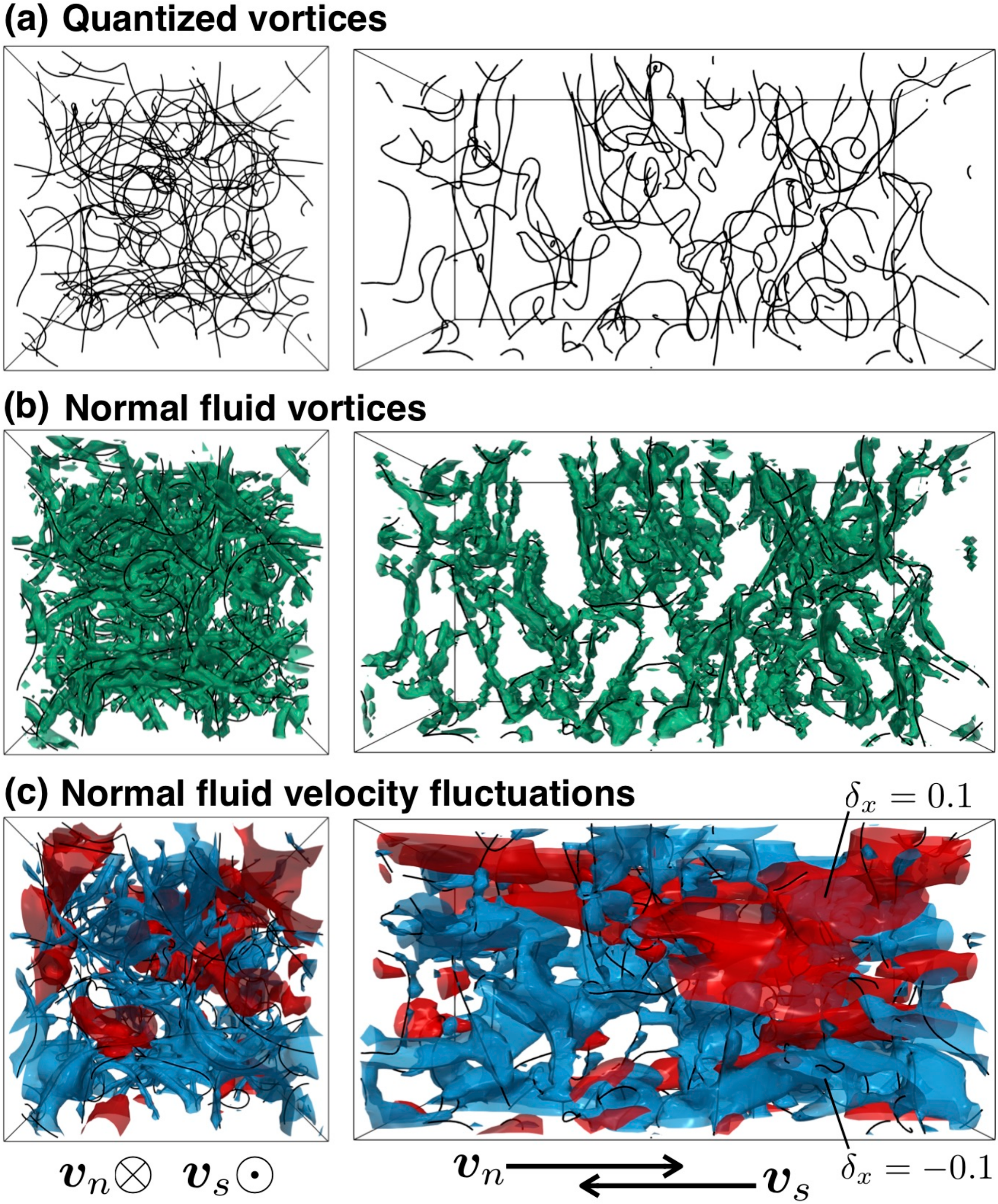}
  \caption
  {
  Three-dimensional structures at $V_n = 2.5 ~\mathrm{mm/s}$.
  (a) Quantized vortices.
  The black lines denote vortex filaments.
  (b) Vortices of the normal fluid.
  The green surfaces denote the positive iso-surfaces of $Q$.
  (c) Velocity fluctuations of the normal fluid.
  The red and blue surfaces denote the iso-surfaces of $\delta_x = 0.1$ and $-0.1$, respectively.
  \label{tangles.pdf}
  }
\end{figure}

Figure \ref{tangles.pdf}(a) shows typical snapshots of the structure of the vortex-filament tangle in the steady state at $V_n = 2.5 ~\mathrm{mm/s}$ (The dynamics are seen in the movie of Supplements).
The tangle becomes anisotropic because the MF $\alpha {\bm s}' \times {\bm v}_{ns}$ in Eq. (\ref{eq:filament}) affects the quantized vortices anisotropically in the counterflow \cite{schwarz88,adachi10}.
To analyze the normal-fluid vortices, we calculate the 2nd invariant
$
Q = \frac{1}{2}  \left ( \omega_{ij} \omega_{ij} - S_{ij} S_{ij}  \right)
$
of the velocity gradient tensor employing vorticity tensor $\omega_{ij} = \frac{1}{2} \left ( \partial v_{n,j}/ \partial x_i - \partial v_{n,i} / \partial x_j \right )$ and strain tensor $S_{ij} = \frac{1}{2} \left ( \partial v_{n,j} / \partial x_i + \partial v_{n,i} / \partial x_j \right )$ \cite{hunt88}.
Specifically, $v_{n,i}$ is the $i$th-component of ${\bm v}_n$.
Figure \ref{tangles.pdf}(b) shows the positive iso-surfaces of $Q = 10.0 ~\mathrm{s^{-2}}$, which show vortex tubes with rotational regions.
The normal-fluid vortices are induced near the vortex filaments because the vortex filaments push the normal fluid through the MF ${\bm F}_{ns}$ locally.
The result is qualitatively consistent with the one-ring simulation \cite{kivotides00}.
The normal-fluid vortex structure which is smaller than the mean vortex-line spacing $\ell$ was not examined in the preceding simulation \cite{yui18}.

It is important to investigate the velocity fluctuations in the normal fluid, which are observed in the PTV experiment \cite{mastracci19}.
We define $\delta_x = {(v_{n,x} - V_n)}/{V_n}$ as the streamwise velocity deviation.
Figure \ref{tangles.pdf}(c) shows the iso-surfaces of $\delta_x = 0.1$ (red) and $-0.1$ (blue).
The normal fluid in the red (blue) region is faster (slower) than the mean velocity.
It is noted that the normal fluid is nearly laminar despite fluctuations.
The negative-fluctuation regions with $\delta_x = -0.1$ arise because the vortex filaments push the normal fluid into the superfluid flow direction $-x$ via MF, and normal-fluid velocity fluctuations remain on the trace.
Specifically, the structures of the negative fluctuations appear to reflect the tangle structure of the filaments.
This refers to a normal-fluid wake caused by quantized vortices \cite{mastracci19}.
The positive fluctuations in red can arise from other mechanisms, {\it e.g.}, the back flow due to the constant mean velocity of the normal fluid.
The structure is larger than the mean vortex-line spacing $\ell$.
The most notable aspect is the strong anisotropy of the velocity fluctuations, which is quantitatively investigated in the following sections.

\begin{figure}
  \centering
  \includegraphics[width=1.0\linewidth]{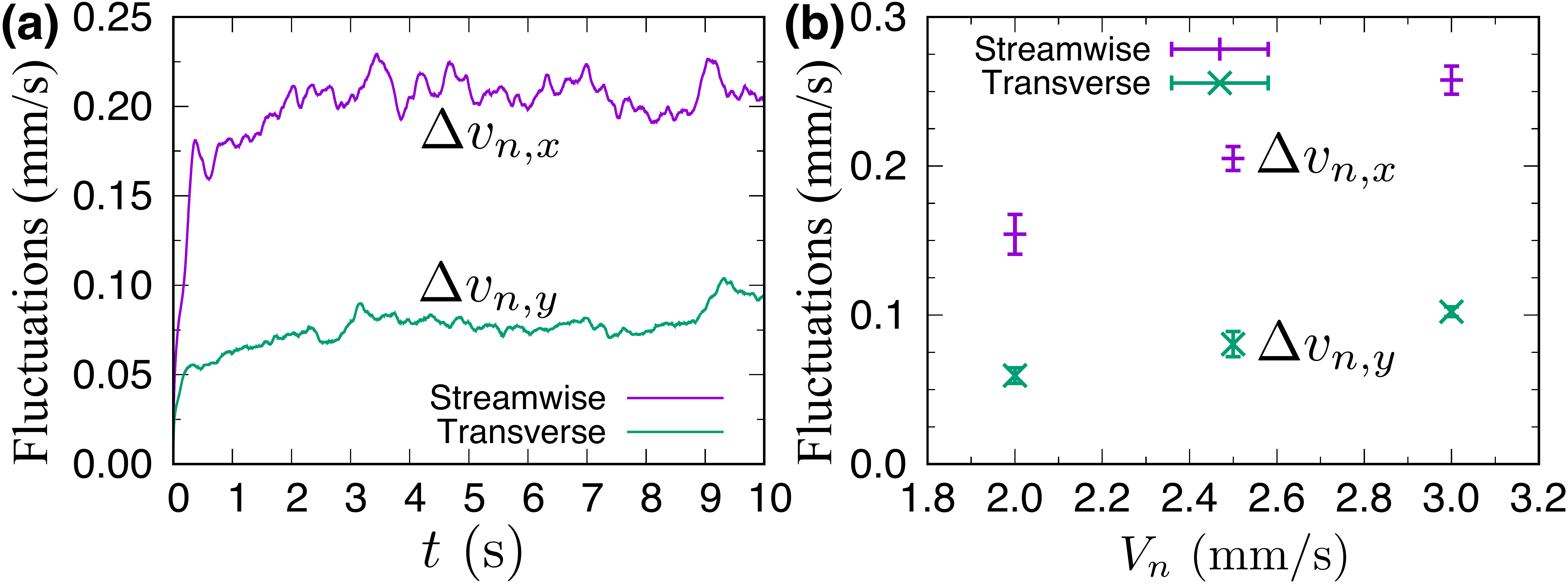}
  \caption
  {
  (a) Velocity fluctuations $\Delta v_{n,x}$ and $\Delta v_{n,y}$ as a function of time at $V_n = 2.5 ~\mathrm{mm/s}$.
  (b) Mean values of the velocity fluctuations as a function of $V_n$.
  \label{fluctuation.pdf}
  }
\end{figure}

As a statistical value of the intensity of the normal-fluid velocity fluctuations, we employ the quantities
\begin{equation}
  \Delta v_{n,x} = \left \langle (v_{n,x} - V_n)^2 \right \rangle^{\frac{1}{2}}, ~~~ 
  \Delta v_{n,y} = \left \langle v_{n,y}^2 \right \rangle^{\frac{1}{2}}.
  \label{eq:fluctuation}
\end{equation}
The value of $\Delta v_{n,x}$ ($\Delta v_{n,y}$) shows the intensity of the velocity fluctuations in the streamwise (transverse) direction.
Figure \ref{fluctuation.pdf}(a) shows the values of $\Delta v_{n,x}$ and $\Delta v_{n,y}$ as a function of time at $V_n = 2.5 ~\mathrm{mm/s}$.
Figure \ref{fluctuation.pdf}(b) shows the values that are temporally averaged over the statistically steady states.
The fluctuations are significantly smaller than the mean flow: $\Delta v_{n,x}, \Delta v_{n,y} \ll V_n$.
Thus, the normal fluid is almost laminar and just disturbed by QT.
The anisotropy of the fluctuations is clearly observed as $\Delta v_{n,x} > \Delta v_{n,y}$, and this anisotropy is a feature of the counterflow QT in contrast to classical turbulence \cite{davidson15}.
The value of $\Delta v_{n,x}$ increases with $V_n$, keeping $\Delta v_{n,x} > \Delta v_{n,y}$.
These results are consistent with the PTV experiments \cite{mastracci18,mastracci19}.
The present values are less than those of the experiments.
This can come from that the MF ${\bm f}$ spreads over the sub-volume.
The smaller sub-volume should reduce the differences between the simulation and the experiment.

\begin{figure}
  \centering
  \includegraphics[width=1.0\linewidth]{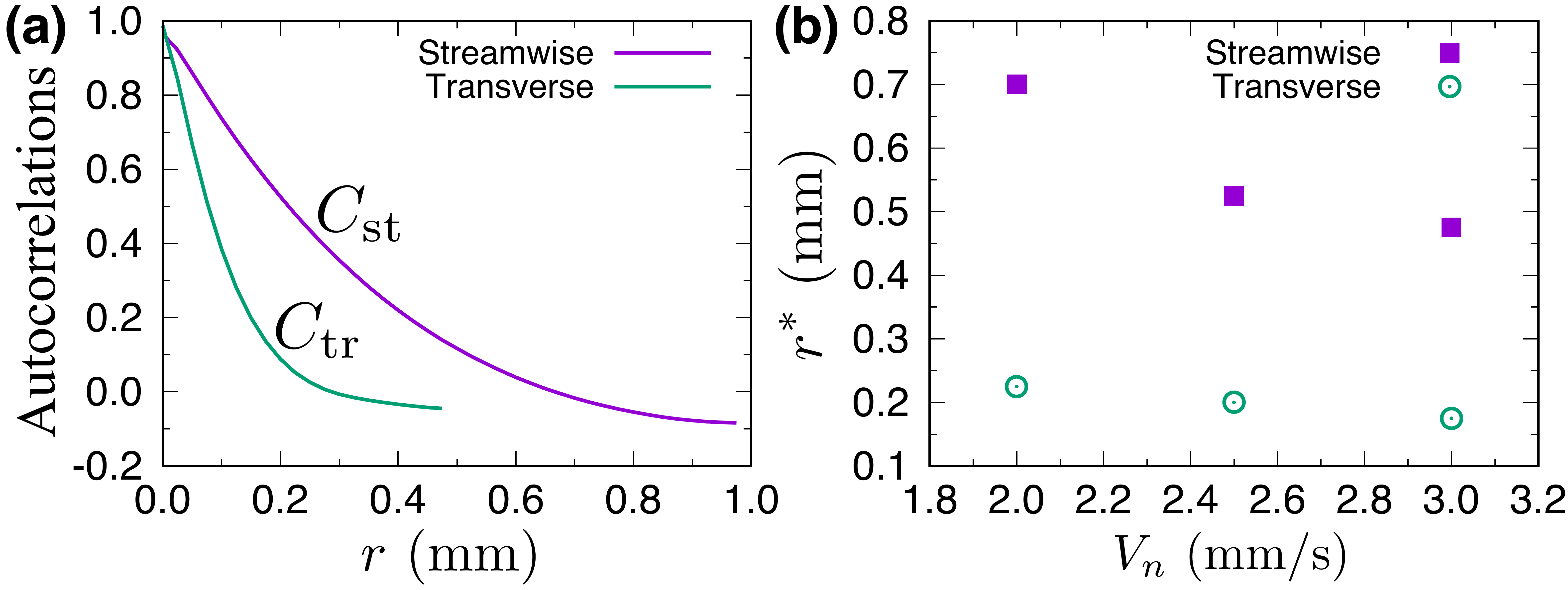}
  \caption
  {
  (a) Autocorrelations as a function of distance $r$ at $V_n = 2.5 ~\mathrm{mm/s}$.
  (b) Distances $r^*$ where the autocorrelations decay to $0.1$.
  \label{correlation.pdf}
  }
\end{figure}

Finally, we investigate the structure of the normal-fluid velocity fluctuations.
We introduce streamwise and transverse autocorrelation functions
\begin{eqnarray}
  C_{\rm st}(r) &=& \frac{ \left \langle \left \langle d_x (x+r,y,z,t) d_x (x,y,z,t) \right \rangle \right \rangle_t }{ \left \langle \Delta v_{n,x}^2 \right \rangle_t}, \\
  C_{\rm tr}(r) &=& \frac{ \left \langle \left \langle d_x (x,y+r,z,t) d_x (x,y,z,t) \right \rangle \right \rangle_t }{ \left \langle \Delta v_{n,x}^2 \right \rangle_t},
\end{eqnarray}
respectively.
Here, $d_x (x,y,z,t) = v_{n,x} (x,y,z,t) - V_n$, and $\langle \cdot \rangle_t$ denotes temporal average.
The widths of the distribution of $C_{\rm st}(r)$ and $C_{\rm tr}(r)$ show the streamwise and transverse sizes of the fluctuation structure, respectively.
Figure \ref{correlation.pdf}(a) shows the values of $C_{\rm st}$ and $C_{\rm tr}$ as a function of distance $r$ at $V_n = 2.5 ~\mathrm{mm/s}$.
Our simulated $C_{\rm st}$ profile, which agrees well with the calculation of a simplified wake-flow model \cite{mastracci19}, differs from the measured velocity autocorrelation at small scales.
This difference may be caused by the uncertainties in the experimental data.
The distances $r^*$ where the autocorrelations decay to $0.1$ are shown in Fig. \ref{correlation.pdf}(b).
Evidently, the streamwise values of $r^{*}$ are significantly larger than the transverse values.
The transverse distances are approximately  $r^* \sim 0.2 ~\mathrm{mm}$, which is comparable to $\ell$.
This agreement is because the fluctuations reflect the structure of the tangle of the vortex filaments, and the fluctuations are localized near the vortex filaments in the transverse direction.
Conversely, the streamwise distances of $r^{*}$ exceed $\ell$.
The streamwise large structures are consistent with the PTV experiment \cite{mastracci19}.
The large structures originate from the normal-fluid wakes caused by quantized vortices as shown in the blue regions of Fig. \ref{tangles.pdf}(c), and also from the positive fluctuations in red.

%%%
{\it Conclusions.---}In the study, we addressed the T1 state by using a numerical simulation of three-dimensional coupled dynamics of the VFM and HVBK equations.
We obtained the laminar normal fluid and turbulent superfluid in statistically steady states, {\it i.e.}, the T1 state.
The normal-fluid vortices were generated near the vortex filaments via MF.
The results indicated that velocity fluctuations of the normal fluid exhibit strong intensity and long-range autocorrelation in the streamwise direction.
Our results are consistent with the PTV experiment \cite{mastracci19}.
This success validates the model and paves the way for future study on the fully coupled dynamics.
The T1-T2 transition could be directly produced with the present method only by increasing the flow velocity.
Moreover, this study is applicable to other important problems such as QT in a realistic solid channel and decaying QT \cite{stalp99,gao16,gao18}.

Elucidating the origin of these velocity fluctuations provides critical insights for some long-standing questions.
For instance, the T1-T2 transition corresponds to a turbulent transition in the normal fluid \cite{tough82}.
%The mechanism that drives this transition is still an outstanding question despite decades of research on counterflow.
The transition mechanism is still an outstanding question despite decades of research on counterflow.
The shear stress from the channel wall, which drives the turbulent transition in classical channel flow \cite{davidson15}, could be responsible for this transition.
But as Melotte and Barenghi pointed out \cite{melotte98}, a new mechanism, {\it i.e.}, the MF, may play a more important role.
This work has identified the velocity fluctuations in the laminar normal fluid, which provides strong support to this view.
These fluctuations may serve as the seed for triggering the normal-fluid turbulent transition \cite{mastracci19}.
It can be naturally confirmed using our model in the future.

%%%
\begin{acknowledgments}
The authors acknowledge W. F. Vinen for insightful discussions.
S. Y. acknowledges the support from Grant-in-Aid for JSPJ Fellow (Grant No. JP19J00967).
H. K. acknowledges the support from the MEXT-Supported Program for the Strategic Research Foundation at Private Universities ``Topological Science'' (Grant No. S1511006) and JSPS KAKENHI (Grant No. JP18K03935).
M. T. acknowledges the support from JSPS KAKENHI (Grant No. JP17K05548).
W. G. acknowledges the support from the National Science Foundation (Grant No. DMR-1807291) and the National High Magnetic Field Laboratory which is supported through the NSF Cooperative Agreement No. DMR-1644779 and the state of Florida.
\end{acknowledgments}

%%%

\setcounter{equation}{0}
\renewcommand{\theequation}{S.\arabic{equation}}
\setcounter{figure}{0}
\renewcommand{\thefigure}{S.\arabic{figure}}

\section{Supplements}
We detail two supplemental materials, namely local coupling of the mutual friction and time dependence of the vortex line density.

%%%
\subsection{Local coupling of mutual friction}
The vorticity ${\bm \omega}_s$ of the superfluid is extremely localized at the vortex core corresponding to $0.1 ~\mathrm{nm}$.
Mutual friction is the interaction between the vortex core of the superfluid and normal-fluid elementary excitations.
Thus, interaction is localized at the vortex filaments, and the normal fluid is affected by the mutual friction at the position of the vortex filaments.
The mutual friction per unit length of the filaments is \cite{barenghi83,schwarz85}
\begin{equation}
  \frac{{\bm f}(\xi)}{\rho_s \kappa} = \alpha {\bm s}' \times ({\bm s}' \times {\bm v}_{ns}) + \alpha' {\bm s}' \times {\bm v}_{ns}.
\end{equation}
The normal fluid is affected by ${\bm f}$ such as the Dirac delta function.

\begin{figure}
  \centering
  \includegraphics[width=0.8\linewidth]{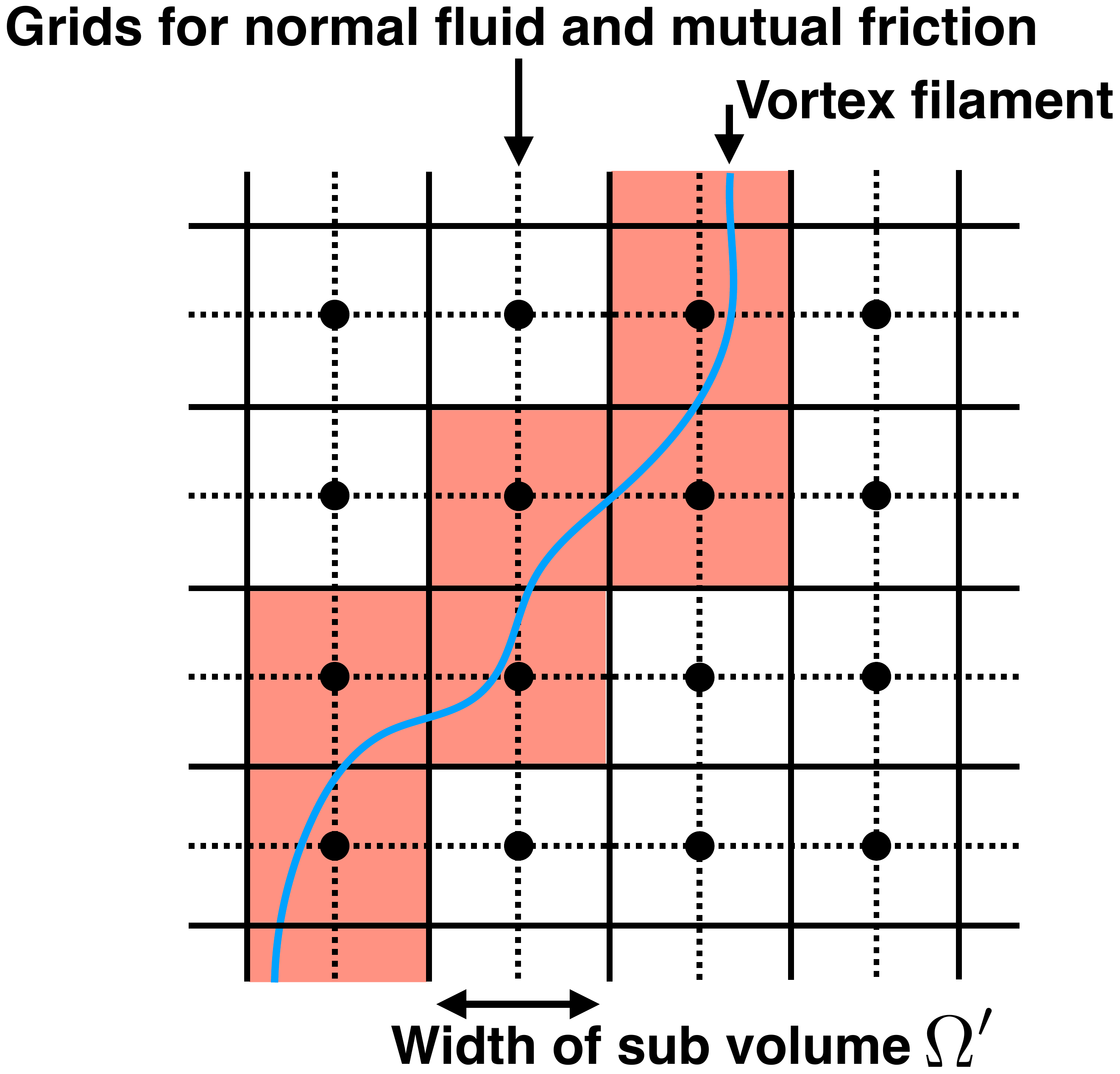}
  \caption
  {
  Schematics of local coupling condition of mutual friction.
  \label{mutual_friction.pdf}
  }
\end{figure}

However, in the numerical simulation of the normal fluid, it is not possible to deal with the delta function behavior because it necessitates infinitely fine resolutions.
Subsequently, we prepare local sub-volumes $\Omega'$, which are obtained by dividing the whole volume $\Omega$.
The mutual friction ${\bm f}$ is averaged in the sub volume.
The expression is as follows:
\begin{equation}
  {\bm F}_{ns} ({\bm r}) = \frac{1}{\Omega' ({\bm r})} \int_{{\mathcal L}' ({\bm r})} {\bm f} ({\xi}) d\xi,
  \label{eqs:friction}
\end{equation}
where the path ${\mathcal L}' ({\bm r})$ shows the vortex filaments in ${\Omega}' ({\bm r})$.
This expression is applicable for numerical simulations.
The selection of the size of $\Omega '$ determines the coupling scale of the mutual friction.
When $\Omega ' = \Delta x \Delta y \Delta z$, in terms of the mean vortex-line spacing $\ell \sim L^{-\frac{1}{2}}$, the local coupling condition is
\begin{equation}
  \ell \gg \Delta x, \Delta y, {\rm and}~ \Delta z,
  \label{eqs:local}
\end{equation}
and the coarse-graining condition is
\begin{equation}
  \ell \ll \Delta x, \Delta y, {\rm and}~ \Delta z.
  \label{eqs:coarse}
\end{equation}
When the mean spacing $\ell$ of the vortex filaments exceeds the widths of $\Omega'$, the mutual friction ${\bm f}$ is not coarse-grained.
In the local coupling condition of Eq. (\ref{eqs:local}), mutual friction only affects the normal fluid at the position of the vortex filaments.
Conversely, in the condition of Eq. (\ref{eqs:coarse}), the sub-volume can contain several vortex filaments.
Thus, the mutual friction ${\bm f}$ is coarse-grained in this condition.
The coarse-graining condition of Eq. (\ref{eqs:coarse}) is useful in examining the macroscopic flow structure of the normal fluid, {\it e.g.}, the dynamics of the laminar normal fluid \cite{yui18}.

The local coupling condition is shown in Fig. \ref{mutual_friction.pdf}.
The dashed lines denote the grids for the normal fluid and mutual friction ${\bm F}_{ns}$.
The solid lines denote the boundary of the local sub-volume $\Omega'({\bm r})$, and the integral in Eq. (\ref{eqs:friction}) is performed along the filaments in the sub-volume.
The sub-volume $\Omega' ({\bm r})$ containing the vortex filaments is indicated in red.
The value of ${\bm F}_{ns}$ is defined at the grid points of the dashed lines.
In the non-colored sub-volumes without vortex filaments, the mutual friction does not work on the normal fluid.
Namely, the mutual friction affects the normal fluid only at the position of the vortex filaments, and this corresponds to the local coupling condition.

\begin{figure}
  \centering
  \includegraphics[width=0.9\linewidth]{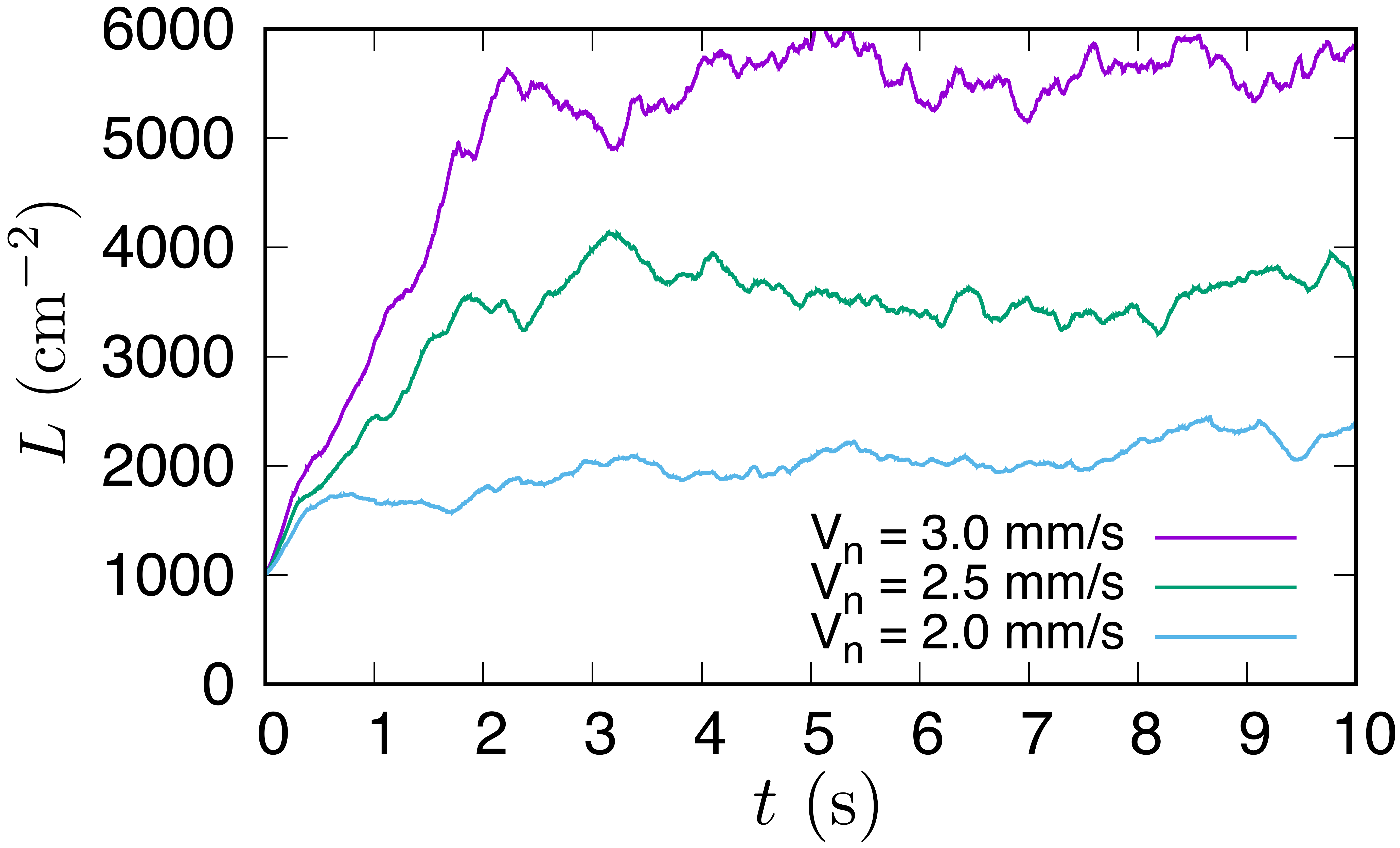}
  \caption
  {
  Vortex line density as a function of time at different values of the mean velocity $V_n$ of the normal fluid.
  \label{density.pdf}
  }
\end{figure}

%%%
\subsection{Vortex line density}
Figure \ref{density.pdf} shows the vortex line density $L$ as a function of time.
The values increase from the initial value and are saturated after a few seconds.
This implies that the generation and dissipation of the vortex filaments compete during those states.
This corresponds to the statistically steady state of QT.
The dynamics of the vortex line density is given by the Vinen's equation:
\begin{equation}
  \frac{dL}{dt} = \chi_1 \alpha V_{ns} L^{\frac{3}{2}} - \chi_2 \frac{\kappa}{2\pi} L^2,
  \label{eqs:vinen}
\end{equation}
where $\chi_1$ and $\chi_2$ are the temperature-dependent parameters \cite{vinen57c}.
The first term on the right-hand side shows the production by the mean counterflow velocity $V_{ns}$.
The second term on the right-hand side corresponds to the decay term.
The two terms are balanced in a statistically steady state.
The steady-state solution of Eq. (\ref{eqs:vinen}) is as follows:
\begin{equation}
  L^{\frac{1}{2}} = \gamma V_{ns},
  \label{eqs:relation_0}
\end{equation}
where
\begin{equation}
  \gamma = \frac{2\pi \alpha \chi_1}{\kappa \chi_2}.
\end{equation}
Specifically, the relation needs another parameter $V_0$ and is expressed as follows:
\begin{equation}
  L^{\frac{1}{2}} = \gamma (V_{ns} - V_0).
  \label{eqs:relation_01}
\end{equation}
This relation is confirmed in the experiments \cite{tough82}.
Our results satisfy the relation in Eq. (\ref{eqs:relation_01}).

%%%

\bibliography{biblio.bib}
\bibliographystyle{apsrev4-2}

\end{document}